%% file: 00_main.tex
\documentclass[10pt,journal,compsoc]{IEEEtran}

\ifCLASSOPTIONcompsoc
  \usepackage[nocompress]{cite}
\else
  \usepackage{cite}
\fi

\usepackage{amsmath}
\usepackage{algorithmic}
\usepackage{url}
\usepackage{hyperref}
\usepackage{graphicx}
\usepackage{amsfonts}
 \usepackage[table,xcdraw]{xcolor}
\usepackage{multirow}
\usepackage{microtype}

\begin{document}
\title{\huge 
%Initializing a Cross-Parameterization
Obtaining a Canonical Polygonal Schema 
from a Greedy Homotopy Basis with Minimal Mesh Refinement}


\author{Marco~Livesu,~CNR IMATI}% <-this % stops a space

\markboth{Journal of \LaTeX\ Class Files,~Vol.~14, No.~8, August~2015}%
{Shell \MakeLowercase{\textit{et al.}}: Bare Demo of IEEEtran.cls for Computer Society Journals}

\IEEEtitleabstractindextext{%
\begin{abstract}
Any closed manifold of genus $g$ can be cut open to form a topological disk and then mapped to a regular polygon with $4g$ sides. This construction is called the \emph{canonical polygonal schema} of the manifold, and is a key ingredient for many applications in graphics and engineering, where a parameterization between two shapes with same topology is often needed. 
The sides of the $4g-$gon define on the manifold a system of loops, which all intersect at a single point and are disjoint elsewhere. 
%In the discrete setting the manifold is typically represented as a triangle mesh, and 
Computing a shortest system of loops of this kind is NP-hard. A computationally tractable alternative consists in computing a set of shortest loops that are not fully disjoint in polynomial time using the greedy homotopy basis algorithm proposed by Erickson and Whittlesey~\cite{erickson2005greedy}, and then detach them in post processing via mesh refinement. %and for high genus shapes bundles of loops often travel along the same chain of mesh edges, 
%thus preventing the realization of a canonical schema. A common workaround consists in refining the mesh to detach them in post processing. 
Despite this operation is conceptually simple, known refinement strategies do not scale well for high genus shapes, triggering a mesh growth that may exceed the amount of memory available in modern computers, leading to failures. In this paper we study various local refinement operators to detach cycles in a system of loops, and show that there are important differences between them, both in terms of mesh complexity and preservation of the original surface. We ultimately propose two novel refinement approaches: the former minimizes the number of new elements in the mesh, possibly at the cost of a deviation from the input geometry. The latter allows to trade mesh complexity for geometric accuracy, bounding deviation from the input surface. Both strategies are trivial to implement, and experiments confirm that they allow to realize canonical polygonal schemas even for extremely high genus shapes where previous methods fail.
%some strategies perform remarkably better than others, bounding the growth in size of the mesh, without significantly deviating from the original surface.
%and ultimately making the canonical polygonal scheme not only a strong theoretical result, but also a suitable practical solution for cross parameterization between shapes.
\end{abstract}

\begin{IEEEkeywords}
Topology, polygonal schema, cut graph, homology, homotopy, cross parameterization
\end{IEEEkeywords}}

\maketitle
\IEEEdisplaynontitleabstractindextext
\IEEEpeerreviewmaketitle

\input{99_utilities}
\input{01_intro}
\input{02_related}
\input{03_method}
\input{04_results}
\input{05_conclusions}

%\appendices

\ifCLASSOPTIONcompsoc
%\else
%  % regular IEEE prefers the singular form
%\fi


\ifCLASSOPTIONcaptionsoff
  \newpage
\fi


\bibliographystyle{IEEEtran}
\bibliography{00_main}

%\begin{IEEEbiography}{Marco Livesu}
%\end{IEEEbiography}

%\vfill


\end{document}

%% file: 99_utilities.tex
%%%%%%%%%%%%
% USEFUL COMMANDS
%%%%%%%%%%%%

\newcommand{\cino} [1]{{\color{magenta}	Cino: #1}}
\newcommand{\com} [1]{{}} % comment stuff out....
\newcommand{\edit} [1]{{\color{red}			#1}} % mark edits in red (e.g. after a revision)

\newcommand{\M}{M} % manifold
\renewcommand{\l}{\ell} % loop
\renewcommand{\P}{P} % polygon
\newcommand{\Q}{Q} % queue
\newcommand{\genus}[1]{{ g( #1 ) }}% genus

%% file: 01_intro.tex
\IEEEraisesectionheading{\section{Introduction}}
\label{sec:introduction}

Generating a cross parameterization between two 3D shapes with same genus is an interesting topological problem with practical impact in many fields. Maps of this kind allow to transfer a signal from one shape to the other, and are exploited by many tools in computer graphics, engineering and medicine for various applications, such as texture mapping, remeshing and shape registration, to name a few. 

\begin{figure}
	\centering
	\includegraphics[width=\linewidth]{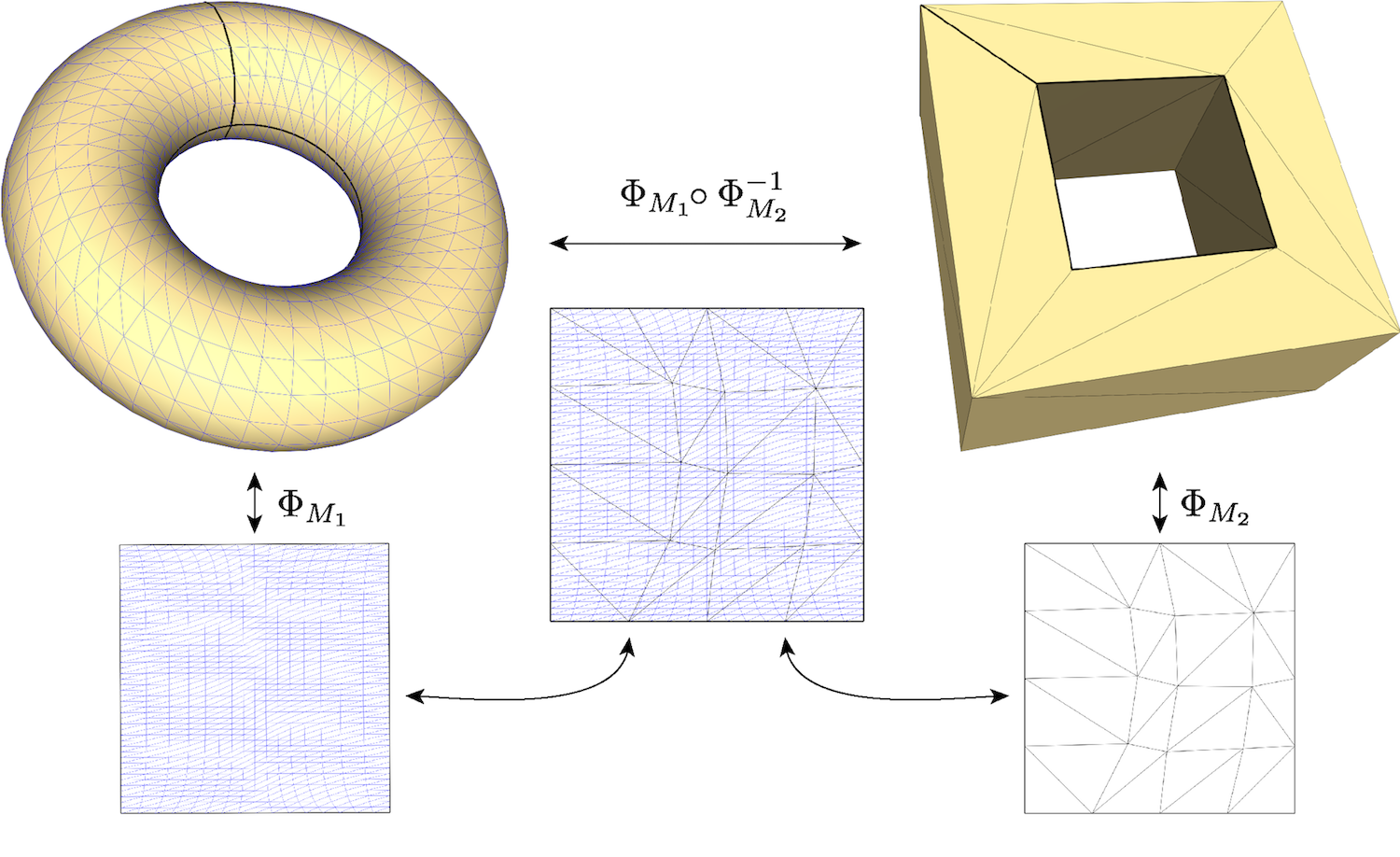}
	\caption{A cross parameterization between a torus and its polycube (computed with~\cite{LVSGS13}). Maps between any two homotopic shapes can be obtained by firstly projecting each shape to its canonical polygonal schema (left,right), and then using it as a medium to travel from one shape to the other (middle). Note that there are $4g$ possible ways to overlap the canonical polygons of two manifolds with genus $g$.}
	\label{fig:pcmap}
\end{figure}

Given two manifolds, a robust way to obtain a cross parameterization consists in cutting both shapes open to topological disks, flatten them on the plane, and then overlap the two disks to obtain a point-to-point correspondence (Figure~\ref{fig:pcmap}). Computational topology 
%guarantees that a homeomorphism between two manifolds with same genus $g$ always exists, and 
provides a sound theoretical framework to perform each of these operations. 

\begin{figure*}
	\centering
	\includegraphics[width=\linewidth]{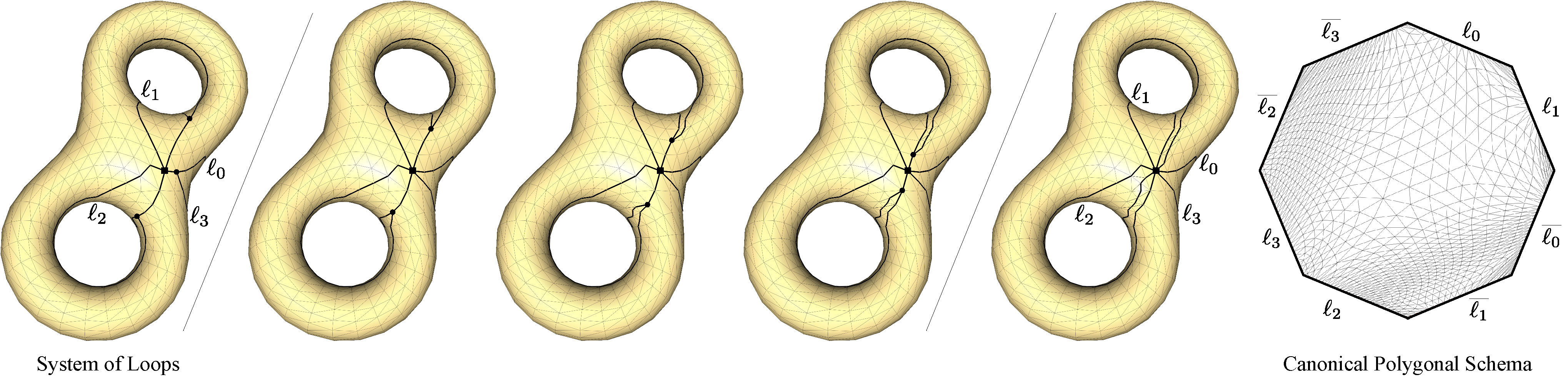}
	\caption{Left: the greedy homotopy basis algorithm generates a system where loops are not fully disjoint ($\l_0$ and $\l_1$ merge at the black circle on top, $\l_0$ and $\l_3$ merge at the black circle in the middle, $\l_2$ and $\l_3$ merge at the black circle at the bottom). Middle: merging points are iteratively pushed towards the origin of the basis (black square) until they all vanish to it. Right: the associated canonical polygonal schema.}% To minimize geometric distortion interior vertices are mapped with Tutte~\cite{tutte1963draw} with cotangent weights~\cite{pinkall1993computing}.}
	%		The cut graph is composed of 4 disjoint loops that emanate from a single source, yielding an octagon. }
	\label{fig:cps}
\end{figure*}

Any closed orientable surface with genus $g$ has exactly $2g$ classes of homotopically independent loops. A system of loops
%$$
%\mathcal{B} = \left\lbrace \ell_0, \ell_1, \dots, \ell_{2g}  \right\rbrace
%$$
containing one loop from each such class is also a homotopy basis~\cite{erickson2005greedy}. If cut along its homotopy basis,
%$\mathcal{B}$, 
the surface becomes a topological disk, hence it can be flattened to the plane. In particular, if all loops emanate from a single source and are disjoint elsewhere, cutting the surface yields a polygon with $4g$ sides, called the \emph{canonical polygonal schema} of the surface~\cite{vegter1990computational} (Figure~\ref{fig:cps}, right). This construction
is a topological invariant,
%depends solely on the topology of the shape, 
 hence any two shapes with same genus share the same polygonal schema, which can then be used as a medium to initialize a cross parameterization between them~\cite{li2008globally,carner2005topology,wang2007polycube}.
%Various algorithms already exploit this construction for practical applications~\cite{li2008globally,carner2005topology,wang2007polycube}.
%Since two polygons with $n$ sides can be overlapped in $n$ different ways, this mapping is not unique. Various heuristics can be used to fix this rotational degree of freedom (e.g. using landmarks).
%One way to compute it consists in mapping both manifolds on their \emph{canonical polygonal schema}, which is a regular 2D polygon having exactly $4g$ sides~\cite{}. All manifolds with same genus share the same polygonal schema, which can then overlapped up to a rotational degree of freedom. 
For this map to be practically useful the cut graph should be shortest, but this latter condition makes the problem NP-hard. A practical  alternative consists in computing in polynomial time a shortest system of loops that possibly overlap at some mesh edge, using the greedy homotopy basis algorithm proposed in~\cite{erickson2005greedy}, and then detach such loops in post processing via mesh refinement. 
Despite apparently trivial, this refinement operation hides some difficulty. In fact, for high genus shapes the system will contain a big number of loops, which will largely snap to the same chains of edges, requiring massive mesh refinement to fully detach them. 

%To minimize distortion, loops should be as straight as possible (hence shortest). To this end, the greedy homotopy basis algorithm proposed in~\cite{erickson2005greedy} imposed as a standard de facto to compute systems of cutting loops, due to its robustness, ease of implementation, and its ability to efficiently compute shortest homotopy basis.% centered at a given point in $O(n \log n)$, and -- by testing each point in the mesh -- the globally shortest homotopy basis in $O(n^2 \log n)$.
%In general, systems of loops computed with~\cite{erickson2005greedy} do not permit the realization of a canonical polygonal schema. This is due to the fact that loops are defined as shortest paths computed directly on the discrete mesh with Dijkstra.
%shortest paths are computed directly in the discrete mesh 
%In this short paper we address a technical issue of the greedy homotopy basis algorithm, which prevents it to create systems of loops that admit a canonical polygonal schema. This is due to the fact that loops are defined as closed chains of mesh edges, and 
%As a result, their structure is limited by mesh connectivity. 

To better understand this observation, we recall that the $2g$ loops of a discrete manifold with genus $g$ should all intersect at a unique mesh vertex. For these loops to be fully disjoint, such vertex should have at least $4g$ incident chains of edges. For example, the Nasty Cheese model shown at the top of Figure~\ref{fig:mosaic} has genus 133, which means that the origin of its system of loops should be located at a mesh vertex with at least 532 neighbors. Having a mesh with such a connectivity is practically impossible. In fact, it is known that the average vertex valence for triangle meshes is equal to $6$, which means that already for a manifold with genus 2 the chances that all loops in the basis will be fully disjoint are tiny, and mesh refinement is necessary. 
%In practice, homotopy bases produced with~\cite{erickson2005greedy} are typically composed of loops that snap to the same chains of mesh edges, and eventually reach the origin of the system with a handful of curves, each one consisting of a bundle of loops that travel along the same discrete path (Figure~\ref{fig:cps}, left). 

%This issue was already observed by 
Li and colleagues~\cite{li2008globally} %. To initialize a cross parameterization they 
proposed to use edge splits 
%mesh refinement 
to detach loops in a greedy homotopy basis. 
%Their refinement strategy is based on the use of the edge split operator. 
In this paper we show that their refinemet scheme does not scale well on high genus shapes. %, and may produce meshes so big that they do not fit the memory of a modern computer.
%Despite technically correct, we observe that their strategy performs poorly for high genus shapes. 
To make a practical example, 
%Let us consider 
the mesh of aforementioned Nasty Cheese originally contained 30K vertices and 60K faces. After refinement %, the mesh has grown considerably, and 
it counts 1.5M vertices and 3M faces. Such an incredible growth (roughly 5K times more vertices and triangles) is impractical for applications, and hinders the applicability of this technique for complex shapes. As we show in Section~\ref{sec:results}, for higher genus shapes the refined mesh may become so big that it does not fit the memory of a modern computer, leading to a failure.

In this short paper we analyze alternative refine strategies to detach loops in a given homotopy basis. We first show that, to locally detach two loops, the edge split operator introduces an amount of new elements that depends on the local complexity of the mesh, whereas the vertex split operator has a constant cost of 2 new triangles and 1 new vertex. We also show that, despite topologically optimal, the vertex split may occasionally require deviation from the input mesh. Based on these considerations we propose two alternative refinement strategies: the first one simply substitutes the edge split with the vertex split, obtaining minimal mesh growth; the second one takes also geometry into account, and tries to use as many vertex splits as possible, switching to the costly edge split only when the former would introduce significative deviation from the reference geometry. Users can easily trade mesh size for geometric fidelity by acting on an intuitive parameter that controls when the switch between these operators should occur.

Experiments confirm that our refinement strategies outperform previous refinement techniques for high genus meshes, thus turning this strong theoretical framework into a practical algorithm to robustly initialize cross parameterizations between shapes of any complexity.

%% file: 02_related.tex
\section{Background and prior works}
\label{sec:background}
Before providing a precise formulation of our problem, we briefly introduce basic notions from computational topology and discuss previous works, also fixing the notation.

A $2-$manifold $\M$ is a topological space where each point is locally homeomorphic to $\mathbb{R}^2$. 
%The genus $\genus{\M}$ is the number of disjoint cycles that can be removed from the manifold without disconnecting it. In other words, the genus counts the handles of the shape.
%Given two manifolds $\M_1,\M_2$, there exists a homeomorphism between them if and only if $\genus{\M_1}=\genus{\M_2}$~\cite{erickson2004optimally}.
In the discrete setting, manifolds are typically represented as triangle meshes. With abuse of notation, in remainder of the paper we will use the symbol $\M$ to denote both the manifold and its combinatorial realization. The interpretation will become evident form the context. 

Any discrete manifold $\M$ can be cut through a subset of its edges to form a topological disk. This set of edges is called the \emph{cut graph} of $\M$, and its nodes and arcs define the points and edges of a 2D polygon, called \emph{polygonal schema} of $\M$~\cite{erickson2004optimally}.
%The $1-$ skeleton $\M_1$ of a discrete manifold $\M$, is the set graph consisting of its vertices and edges. The \emph{cut graph} $G$ of $\M$, is a subset of $\M_1$ such that $\M \setminus G$ is a topological disk. Topologists call the disk $\M \setminus G$ a \emph{polygonal schema} of $\M$.
The \emph{canonical} polygonal schema is a mapping of $\M$ to a regular polygon with $4g$ sides, where $g$ is the genus of $\M$. The cut graph associated to such a schema designs on $\M$ a system of $2g$ loops $L = \left\lbrace \l_0, \l_1, \dots, \l_{2g} \right\rbrace$ that are fully disjoint except at a common vertex, called the \emph{origin} (or \emph{root}) of the system. The corners of the $4g-$gon are the images of the origin, and the edges are images of the loops, which are ordered according to the gluing scheme
$$ \l_0,\l_1, \overline{\l_0}, \overline{\l_1}, \dots, \l_{2g-1},\l_{2g}, \overline{\l_{2g-1}}, \overline{\l_{2g}} \: ,$$
with $\l_i$ and $\overline{\l_i}$ being two copies of a loop $l_i \in L$ (Figure~\ref{fig:cps}). 
The canonical polygonal schema has two fundamental properties:
\begin{itemize}
\item it is a topological invariant, meaning that two manifolds with same genus map to the same polygon (up to a rotational degree of freedom);
\item it is optimal, in the sense that among all the possible polygonal schemas, the canonical polygon has the least number of edges (i.e. there exists no $k-$gon with $k<4g$ that is the cut graph of a manifold $M$ with genus $g$~\cite{dey1995new})
\end{itemize}

Polygonal schemas play a central role in computer graphics, where they 
%are heavily used to realize the so called $uv$ maps, that is, maps of a surface embedded in $R^3$ on the parametric plane. These maps 
are at the basis of numerous applications, such as texture mapping~\cite{sheffer2005abf++}, remeshing~\cite{alliez2002interactive}, compression~\cite{taubin1998geometric}, and morphing~\cite{kraevoy2004cross}, to name a few. In particular, the properties of the canonical schema make it an appealing starting point to initialize a mapping between two shapes with same genus~\cite{li2008globally,carner2005topology,wang2007polycube}. In fact, as shown in Figure~\ref{fig:pcmap}, given two manifolds $\M_1,\M_2$ with genus $g$, and denoting with $\Phi_{\M_1}$ and $\Phi_{\M_2}$ their one-to-one maps to the canonical polygon $\P_{4g}:$
$$
\begin{array}{c}
\Phi_{\M_1} : \M_1 \leftrightarrow \P_{4g}\\
\Phi_{\M_2} : \M_2 \leftrightarrow \P_{4g}
\end{array}\:,
$$
a cross parameterization $\Phi : \M_1 \leftrightarrow \M_2$ can be obtained through the composition
$$\Phi = \Phi_{\M_1} \circ \Phi_{\M_2}^{-1}\:.$$

Topologists and pratictioners in computer graphics have widely investigated the problem of computing cut graphs for discrete manifolds. Typically, the goal is to find the cut graph with minimal length, or the one that contains the least number of edges. Erickson and Har-Peled showed that both problems are NP-hard, and proposed a greedy algorithm to compute a $O(\log^2 g)-$approximation of the minimum cut graph in $O(g^2 n \log n)$~\cite{erickson2004optimally}. The so generated cut graphs are not necessarily canonical, hence are not suitable for cross parameterization. Dey and Shipper~\cite{dey1995new} propose a linear time method to compute a polygonal schema using a breadth-first search on the dual graph. Their cut-graph is not guaranteed to be shortest, and may not yield the canonical schema as well. In~\cite{de2005optimal} Colin de Verdi{\`e}re and Lazarus propose a polynomial time algorithm that inputs a system of loops, and shrinks it in order to find the shortest system of loops in the same homotopy class. To mimic the continuous framework, the authors \emph{"allow the loops to share edges and vertices in the mesh, provided that they can be spread apart on the surface with a thin space so that they become simple and disjoint except at the origin"}. The authors do not explain how this operation can be performed, and what impact it has on mesh size. In this paper we focus on this very specific problem, aiming to find the mesh refinement strategy with minimal impact on the input manifold, both in terms of number of discrete elements and geometric fidelity. In~\cite{lazarus2001computing} and~\cite{vegter1990computational} methods to compute system of loops that realize a canonical polygonal schema are presented. As already acknowledged by Lazarus and colleagues in their final remarks \emph{"the obtained loops look too much jaggy and complex to be of any use for practical applications. More work needs to be done in this direction taking into account the geometry of the surface"}~\cite{lazarus2001computing}.
%These methods produce jaggy irregular loops, as they do not take into account the geometry of the surface, hence -- As already acknowledged by Lazarus and colleagues -- have little practical utility.
To this end, a big step forward was done by Erickson and Whittlesey with their greedy homotopy basis algorithm~\cite{erickson2005greedy}. At the time of writing, this can be considered to be the state of the art for computing arbitrary polygonal schemas on discrete manifolds. Their method uses the tree-cotree decomposition~\cite{eppstein2003dynamic}, and is guaranteed to find the shortest system of loops centered at a given mesh vertex in $O(n \log n)$, and -- by testing each point in the mesh -- the globally shortest system of loops in $O(n^2 \log n)$. It is interesting to notice that while the computation of the shortest cut graph is NP-hard, the shortest system of loops is easy to compute. The difference between these entities relies in how lengths are computed: in a cut graph, each edge in the cut counts once; in a system of loops, each edge counts as many times as the number of loops in the system that traverse it. It follows that the systems of loops computed with the greedy homotopy basis algorithm are practically useful (because they are shortest), but do not allow to realize a canonical schema (because multiple loops snap to the same mesh edges), hence cannot be used to initialize a cross parameterization between two manifolds.

The use of mesh refinement to detach loops in a given homotopy basis is mentioned in a few aforementioned papers but, to the best of our knowledge, only~\cite{li2008globally} provided an actual algorithm. A similar method was possibly proposed already in~\cite{vegter1990computational}, but the manuscript misses some technical details, and the algorithm is hardly reproducible. To the best of the author's knowledge, no alternative refinement schemes have ever been proposed in previous literature.

%% file: 03_method.tex
\section{Problem Statement and Overview}
Given a discrete manifold $\M$ with genus $g$, our objective is to generate a cut graph that realizes a canonical polygonal schema of $\M$, enabling a map to a regular $4g-$gon. Our algorithm inputs $\M$ and a system of loops $L = \left\lbrace \l_0, \l_1, \dots, \l_{2g} \right\rbrace$. Loops in $L$ are assumed to all emanate from the same origin $O(L)$, but may not be fully disjoint, thus violating the necessary condition to realize a canonical schema, that is
\begin{equation}
\bigcap_{\l_i \in L} \l_i  = O(L)\:.
\label{eq:nec_cond}
\end{equation}
Our method outputs a refined manifold $\M'$ and a new system of loops $L'$, such that $L'$ satisfies Equation~\ref{eq:nec_cond}, and the refinement of $\M'$ is minimal. Without loss of generality, we assume that the input $L$ is computed with the greedy homotopy basis algorithm~\cite{erickson2005greedy}. This is just a practical choice to ensure that loops are shortest. The method works also if loops are not shortest paths, provided that if at some point two loops merge, they follow the same path until they reach the origin $O(L)$.

\begin{figure*}[h]
	\centering
	\includegraphics[width=\linewidth]{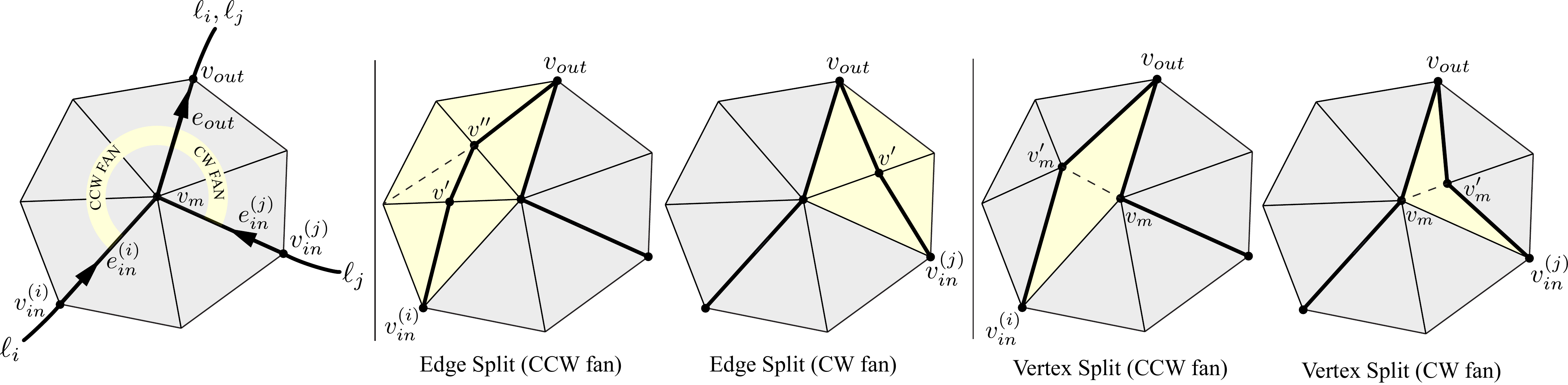}
%	\vspace{-0.4cm}
	\caption{Left: loops $\l_i,\l_j$ meet together at a merging vertex $v_m$. From that point on, they travel together towards the origin of the loop system, $O(L)$. Edges incident to $v_m$ that are traversed by $\l_i,\l_j$ can be locally oriented such that there is one outgoing edge $e_{out}$, traversed by $\l_i,\l_j$, and two ingoing edges, traversed by one loop each (see black arrows). Rotating from $e_{out}$ in both directions until the first ingoing edges are found defines two fans of mesh elements (CW and CWW). 
		%Right: alternative mesh refinements strategies to locally detach $\l_i$ and $\l_j$ around $v_{m}$. 
		Middle: using the edge split to to locally detach $\l_i$ and $\l_j$ around $v_{m}$ using the CWW fan and the CW fan. Right: same result, obtained using the vertex split operator.}
	\label{fig:merge_vert}
\end{figure*}

\subsection{General Algorithm}
\label{sec:overview}
To devise a refinement algorithm we start from a basic observation: loops in the system may be partially overlapping, but can never be entirely coincident. This is ensured by the fact that $L$ is a system of loops in the sense of~\cite{de2005optimal}, hence it is also a cut graph of $\M$. If two loops were coincident, $\M \setminus L$ would not be a topological disk, thus $L$ could not be a cut graph in the first place. It follows that if two loops share a portion of their path towards the origin of the system, there should be a mesh vertex where they begin to coincide. We call this point a \emph{merging} vertex. Figure~\ref{fig:merge_vert} (left) shows an example of merging vertex where two loops collapse into a single discrete path that takes to the origin of the system. 
%As shown in the figure, vertices in the one ring of a merging vertex can be of three types: (i) ingoing vertices ($v_{in}$) , (ii) outgoing vertices ($v_{out}$), and all the other vertices. 
Note that the number of loops incident to a merging vertex can be much higher (for a manifold with genus $g$ the worst case scenario is $2g-1$). Moreover, each incoming path can be either a single loop or a bundle of multiple loops that already joined at a previous merging vertex. From a computational perspective there is no difference between these cases, single loops or bundles of loops can all be locally disjoint using the same refinement operators.
%Edges incident to $v_m$ that are traversed by some loop can always be locally oriented according to the number of loops that contain them. The edge hosting the maximum number of loops is the one that points towards the origin of the system. We call it \emph{outgoing} edge ($e_{out}$), and also call its vertex opposite to $v_m$ outgoing vert ($v_{out}$). All the other edges traversed by some loop are \emph{ingoing} edges ($e_{in}$), and their vertices opposite to $v_{m}$ are ingoing vertices ($v_{in}$). In the general case there can be even more incoming edges, and each one could host a single loop or a bundle of loops that joined at a previous merging vertex and are traveling together towards the origin of the system. In either case, the local processing the merging vertex is the same.

%Given this premise, the general schema of the algorithm is straightforward. 
The main idea of the algorithm is to iteratively push each merging vertex one step forward towards the origin of the system of loops $O(L)$, until all merging points converge to it and Equation~\ref{eq:nec_cond} is satisfied.
%where eventually all such merging vertices will vanish. %, generating a system of loops with no merging vertices other than $O(L)$.
%push all merging vertices forward towards the origin of the system of loops, until they vanish to it. 
In the initialization step, all the merging vertices in $L$ are identified and stored in a queue $\Q$. Then, merging vertices $v_m$ are iteratively extracted and the mesh is locally refined, making sure that all incoming loops traverse the one ring of $v_m$ along a dedicated path. After refinement, the merging point of all such loops has moved to a new mesh vertex which was originally in the one ring of the current $v_m$. If such a point is not the origin of the system of loops, it is added to the queue. The algorithm stops when $\Q$ is empty.
%as follows: the mesh is locally refined around each $v_m$, making sure that its one ring contains as many disjoint paths that converge to $v_{out}$ as the number of ingoing vertices $v_{in}$. Loops in the one ring of $v_m$ are then locally re-assigned to such disjoint paths. As a result, the merging vertex has moved one step towards the origin of the system, and is located at $v_{out}$, which is added to the queue.
%the outgoing vertex $v_{out}$ has now become a new merging vertex, and -- if it is not the origin of the system -- it is enqueued in $\Q$. 
%The algorithm converges when $\Q$ is empty. 
At that point there won't be any merging vertex in $L$ but $O(L)$, thus Equation~\ref{eq:nec_cond} is satisfied, and a canonical polygonal schema of the refined manifold $\M'$ along the newly generated system $L'$ can be computed (Figure~\ref{fig:cps}).
\begin{figure}
	\centering
	\includegraphics[width=.9\columnwidth]{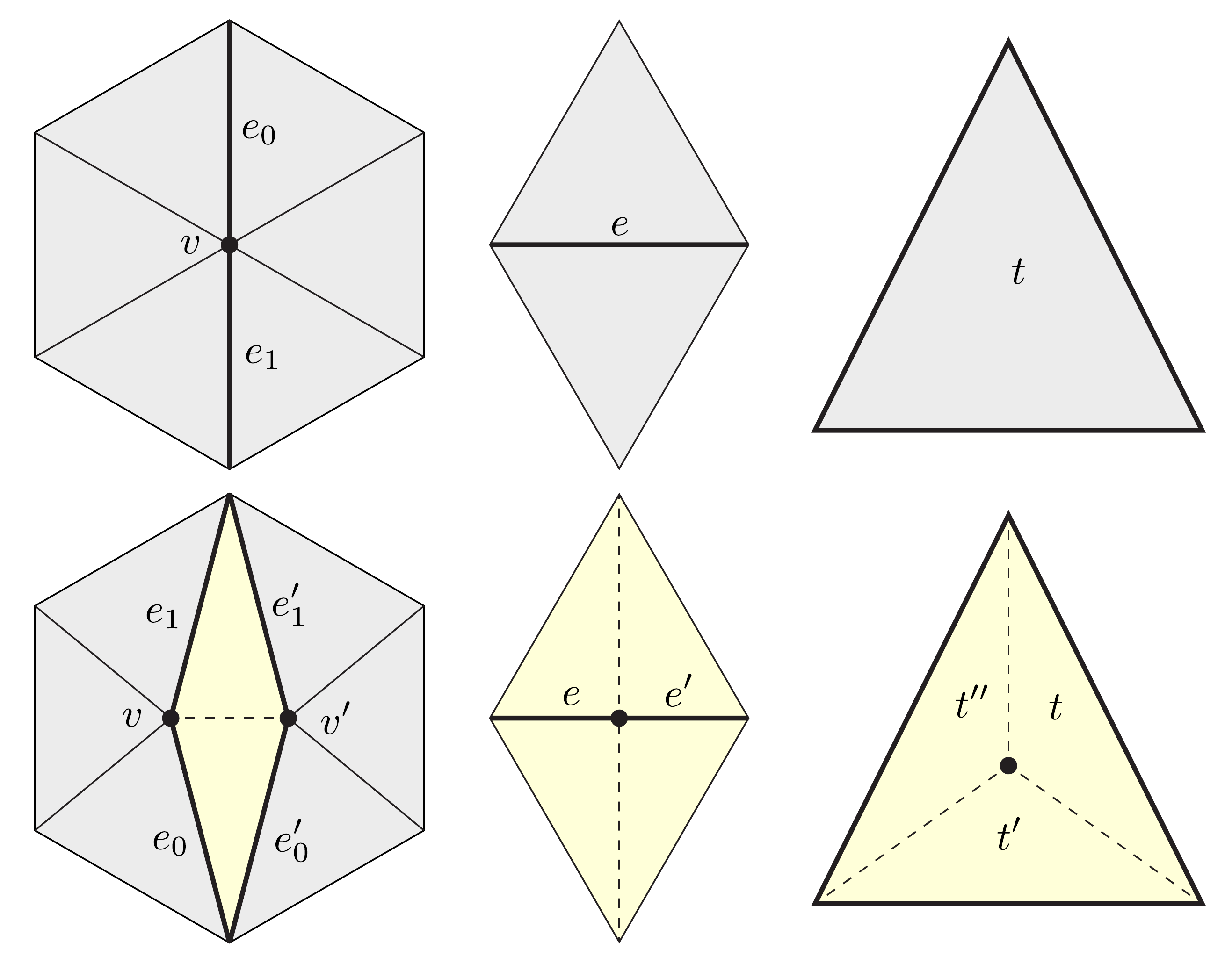}
	\caption{The three possible refinement operators for a triangle mesh. Left: splitting vertex $v$ along its incident edges $e_0,e_1$; middle: splitting edge $e$ at its midpoint; right:  split a triangle $t$ into three sub triangles.}
	\label{fig:split_operators}
\end{figure}
Note that the algorithm above does not provide any detail on how the local refinement is performed. There are several options, which produce different results in terms of number of new elements inserted in the mesh, and geometric distance between $\M$ and $\M'$. In Section~\ref{sec:refinement_analysis} we present all the possible alternatives, discussing pros and cons of each strategy.

\section{Local Refinement Operators}
\label{sec:refinement_analysis}
In this section we explore all the alternative ways to split the elements of a simplicial mesh to detach a set of loops around a merging vertex. The basic ingredients for this operation are illustrated in Figure~\ref{fig:split_operators}. The refinement strategy based on the edge split operator discussed in Section~\ref{sec:edge_split} was already presented in~\cite{li2008globally}. To the best of our knowledge, the alternatives presented in Section~\ref{sec:vert_split} and~\ref{sec:hybrid_refinement} are novel.
%A simplicial complex of dimension 2 can be locally refined by splitting its 0 simplices (i.e. vertices), 1 simplices (i.e. edges) or 2 simplices (i.e. triangles). In this section, we analyze how these operators -- illustrated in Figure~\ref{fig:split_operators} -- can be used to locally detach a set of loops around a merging vertex. 

The typical configuration is the one shown in Figure~\ref{fig:merge_vert}, where two loops, $\l_i,\l_j$ meet at merging vertex and, from that point on, proceed together towards the origin of the system of loops $O(L)$. Edges traversed by some loop can be locally oriented, such that there is one outgoing edge $e_{out}$ that points towards the origin of the system, and two (or more) ingoing edges $v_{in}$, which all converge to the merging vertex. Rotating from the outgoing edge $e_{out}$ clock-wise and counter clock-wise towards the first ingoing edges, defines two ordered fans of mesh elements. These are the two alternative domains that can be used to locally refine the mesh, defining two disjoint paths for $\l_i$ and $\l_j$ within the umbrella of their merging vertex $v_m$. In the following sub-sections we will detail how each splitting operator can be used to perform such operation.

\subsection{Edge Split}
\label{sec:edge_split}
Considering a merging vertex $v_m$ and the ordered fan of edges $E = \left\lbrace e_1, \dots, e_n \right\rbrace$ in between an ingoing edge $e_{in}$ and an outgoing edge $e_{out}$, a unique path connecting the associated ingoing vertex $v_{in}$ and the outgoing vertex $v_{out}$ can be obtained by splitting all edges in $E$. Denoting with $v_i$ the splitting point of edge $e_i$, the path $\left\lbrace v_{in}, v_0,  \dots, v_n, v_{out} \right\rbrace$ is entirely defined within the triangle fan span by $E$, and is also completely disjoint from any other path connecting $v_{in}$ and $v_{out}$. Figure~\ref{fig:merge_vert} (middle) shows its application to the CCW and CW fans of edges around the merging point $v_m$. Note that splitting the CCW edge fan introduces 2 new vertices and 4 new triangles, whereas splitting the CW edge fan introduces 1 new vertex and 2 new triangles. In the general case, the mesh grows linearly with the size of the edge fan, and the growth amounts to $ \vert E \vert$ new vertices, and $2\vert E \vert$ new triangles. Since there are always two alternative edge fans to be split (CW or CCW), to minimize mesh growth it is preferable to always split the fan with smallest size. Note that the edge split requires that $\vert E \vert > 0$. If the fan of elements in between $e_{in},e_{out}$ contains only one triangle and zero edges, loops can be locally split only with the vertex or the triangle split operators.

\subsection{Vertex split}
\label{sec:vert_split}
\begin{figure}
	\centering
	\includegraphics[width=.75\columnwidth]{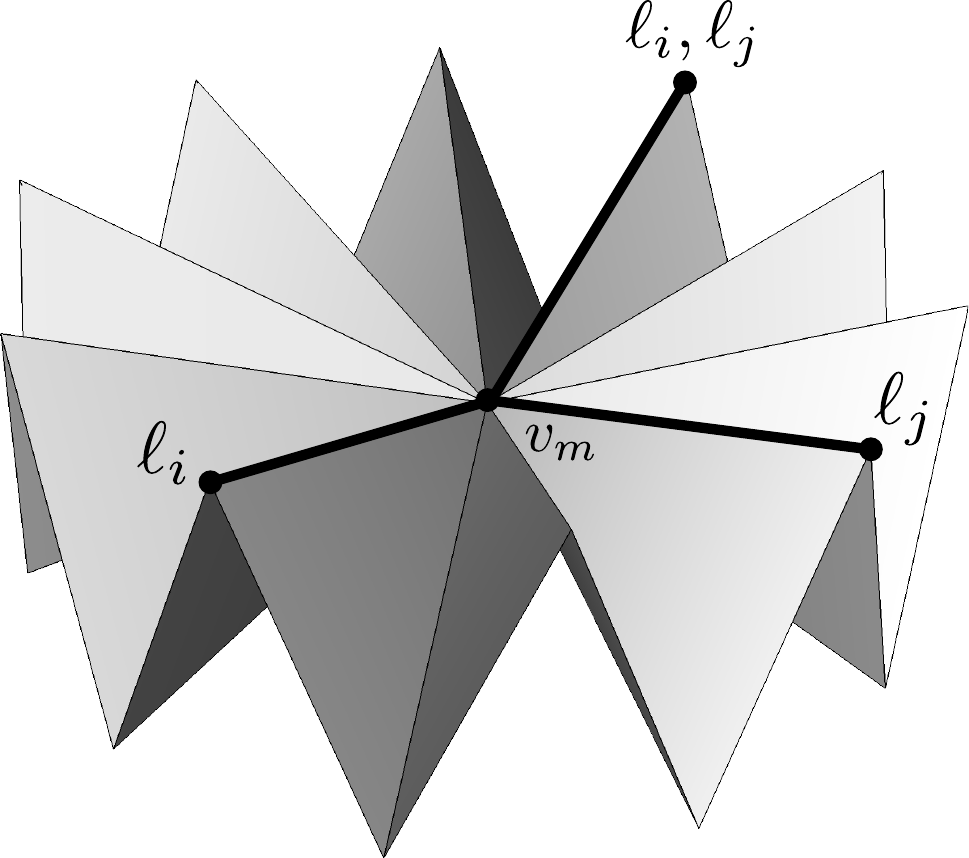}
	\caption{Detaching loops $\l_i,\l_j$ around their merging vertex $v_m$ using the vertex split operator unavoidably deviates from the mesh. Denoting with $v_m'$ the split copy of $v_m$, the two triangles incident to edge $(v_m,v_m')$ deviate from the reference geometry, unless $v_m$ and $v_m'$ coincide. In such case, both triangles will be degenerate.}
	\label{fig:failure_vert_split}
\end{figure}

Considering the same edge fan $E = \left\lbrace e_0, e_1, \dots, e_n \right\rbrace$ around a merging vertex $v_m$, a unique path connecting $v_{in}$ and $v_{out}$ can also be obtained by splitting $v_m$ along the ingoing and outgoing edges $e_{in}, e_{out}$ that bound $E$. Figure~\ref{fig:merge_vert} (right) shows an application of this refinement scheme to the CCW and CW fans around the merging point $v_m$. Note that in both cases the number of new mesh elements amounts to 1 new vertex and 2 new triangles. Differently from the edge split case, this growth is invariant and does not depend on the local complexity of the mesh. 
%This explains why the vertex split operator introduces far less new elements when detaching loops in high genus meshes. 
Although preferable from a topological point of view, the vertex split operator has a geometric limitation: depending on the geometry of the mesh, the two new triangles incident to the new edge $(v_m,v_m')$ will not adhere to the original mesh, introducing a deviation from the target geometry. An example of failure case is depicted in Figure~\ref{fig:failure_vert_split}. In general, any time the fan of triangles span by $E$ is not planar, the vertex split operator introduces such a deviation.

%Considering the same edge fan $E = \left\lbrace e_{in}^{(i)} , e_1, \dots, e_{out} \right\rbrace$, the vertex split operator works by splitting the merging vertex $v_m$ along its incident edges $e_{in}^{(i)}$ and $e_{out}$. As a result, two new triangles and one new vertex, $v_m'$ are added to the mesh (Figure~\ref{}). The new path connecting $v_{in}^{(i)}$ with $v_{out}$ will then be formed by the chain of vertices $\left\lbrace v_{in}^{(i)}, v_m', v_{out} \right\rbrace$. It should become now evident the biggest difference between th edge split and the vertex split operator: while the former introduces a number of new elements in the mesh that scales linearly with the number of elements in the one ring of $v_m$, the vertex split operator has a constant cost of one new vertex and two new triangles, regardless of vertex valences. 

%
\subsection{Triangle split}
\label{sec:triangle_split}
\begin{figure}
	\centering
	\includegraphics[width=.95\columnwidth]{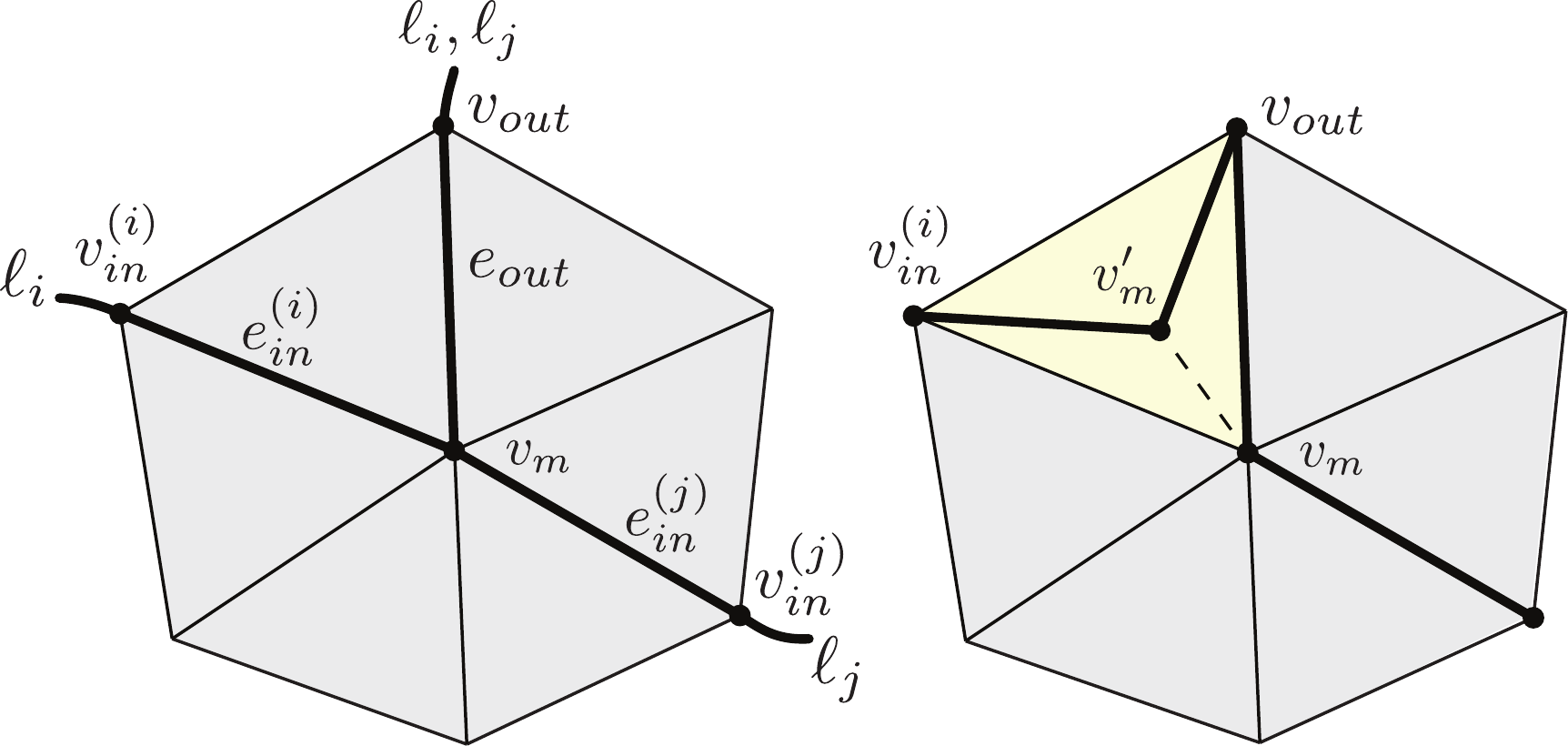}
	\caption{%The CCW fan is composed of a single triangle, shared by edges $e_{in}^{(i)}, e_{out}$. 
		Splitting triangle $v_{in}^{(i)}, v_m, v_{out}$ locally detaches loops $\l_i,\l_j$. Note that $\l_i$ is not a shortest loop, because $\vert v_{in}^{(i)} - v_{out} \vert < \vert v_{in}^{(i)} - v_{m} \vert  + \vert v_{m} - v_{out} \vert$. Also note that the triangle split is conceptually equivalent to splitting vertex $v_m$ along the edges $e_{in}^{(i)}, e_{out}$.} 
	\label{fig:triangle_split}
\end{figure}

Differently form the edge split and the vertex split operator, the triangle split operator can be used to locally detach a pair of loops if and only if the ingoing and the outgoing edges share the same triangle. In that case, adding a new vertex inside the triangle and connecting it to the three corners with new edges generates an alternative path from the ingoing vertex $v_{in}$ and the outgoing vertex $v_{out}$, without passing from the merging vertex $v_m$ (Figure~\ref{fig:triangle_split}). Note that this operation is equivalent to performing a vertex split of $v_m$ along the edges $e_{in},e_{out}$. Also note that if the input system of loops is shortest -- as in the case of~\cite{erickson2005greedy} -- this configuration will never occur. If fact, due to the triangular inequality 
$$
\vert v_{in} - v_{out} \vert < \vert v_{in} - v_{m} \vert  + \vert v_{m} - v_{out} \vert
$$
the path $\left\lbrace v_{in}, v_{out} \right\rbrace$ will always be shorter than the path $\left\lbrace v_{in} ,v_{m} , v_{out} \right\rbrace$, hence $v_m$ would not be a merging vertex. Considering its limited applicability and the fact that, even when usable, the triangle split is equivalent to the vertex split, this is not a suitable operator to detach loops in a cut graph.

\subsection{Hybrid split}
\label{sec:hybrid_refinement}
The analysis of standard refinement operators revealed that:
\begin{itemize}
\item the edge split operator can always be used to locally detach loops around a merging vertex without deviating from the input geometry, and introduces an amount of new mesh elements that scales linearly with the number of elements incident to the merging vertex;
\item the vertex split operator can always be used to locally detach loops around a merging vertex, at the fixed cost of one new vertex and two new triangles. Despite optimal from a topological standpoint, this strategy has a geometric limitation: if none of the two triangle fans are coplanar, the new triangles deviate from the original surface;
\item the triangle split operator can be used to locally detach loops around a merging vertex if and only if one of the two triangle fans is composed of a single element. When applicable, it is equivalent to the vertex split operator.
\end{itemize}

We introduce here a fourth option, which aims at combining the positive aspects of the first two operators, using as many vertex splits as possible to minimize mesh growth, and switching to the costly edge split to avoid deviation from the surface. The method is extremely simple, and seamlessly integrates in the global detaching algorithm described in Section~\ref{sec:overview}. 

Given a merging vertex, the hybrid local refinement first checks whether either the CW or CCW fans of elements aside the outgoing edge are roughly planar. If so, it splits the merging vertex along the ingoing and outgoing edges that bound such fan, as described in Section~\ref{sec:vert_split}. If none of the fans are roughly planar, it locally refines the mesh using the edge split operator as described in Section~\ref{sec:edge_split}. 
To measure planarity we simply consider the maximum angle between the normals $\textbf{n}_i,\textbf{n}_j$ of two triangles $i,j$ in the fan of triangles
\begin{equation}
\arg \max_{i,j} \angle (\textbf{n}_i , \textbf{n}_j)
\label{eq:planarity}
\end{equation}
If the angle above evaluates zero, the fan is planar and the vertex split operator can be used without introducing any deviation from the input surface. In all other cases some deviation from the reference geometry will occur. Assuming the mesh is planar and the vertex split operator is used, positioning the new vertex $v_m'$ in the one ring of the merging vertex $v_m$ is also critical to ensure that no triangle will flip its orientation in the refined mesh. Making sure that $v_m'$ stays inside the polygon defined by the boundaries of the edge fan is not enough, because such polygon may be non convex ($e_{in}$ and $e_{out}$ may form a concave angle). We practically solve this issue by 
%Assuming the mesh is locally planar and the vertex split operator is used, a good location for the new copy of the merging vertex must be found. To this end, note that the boundaries of the edge fan define a polygon that may be non convex, because $e_{in}$ and $e_{out}$ may form a concave angle (measured from the side of the fan of elements being split). It follows that, even if the new point is inside the fan, depending on its position some of the new triangles may be flipped. In our implementation, we 
initializing the new point as
$$
v_m' =  (1-\lambda) v_m +  \lambda v_e
$$
where $\lambda$ is initially set to 0.75, and $v_e$ is the vertex opposite to $v_m$ along the edge $e$, which is median in the edge fan being split. If any of the triangles incident to $v_m'$ is flipped, we halve $\lambda$ and update its position, until a valid position is found. Such a position always exists if $e_{in},e_{out}$ do not coincide. In practice we use the vertex split operator even when the fan of elements is roughly planar. To do so we simply test Equation~\ref{eq:planarity} with a threshold angle set by the user. For more details on the actual values refer to Section~\ref{sec:results}.

%Since in many practical applications a tiny deviation from the mesh is not harmful, we allow the user to balance mesh growth with geometric fidelity by thresholding the result of Equation~\ref{eq:planarity}. In all our experiments we used a fixed threshold of 5 degrees, above which the edge split operator is used instead of the cheaper vertex split. 

%Having perfectly planar mesh regions may not seem a common scenario;  however, in many practical applications one of the meshes in a cross parameterization is a parametric space made of vast planar regions (e.g. a polycube~\cite{LVSGS13}).
%, and is largely planar regions abound
%with a strong structure where one of the two manifolds is a parametric domain (e.g. a polycube~\cite{LVSGS13}) vast regions of the mesh will be planar and could benefit a lot from using our hybrid refinement scheme. 
%This often happens in many applicative scenarios, e.g. when the manifold represents a parametric domain made of vast planar regions~\cite{LVSGS13}. 
%Conversely, for discrete manifolds that represent samplings a smooth surface, the triangles around a merging vertex will hardly be perfectly planar, and a  deviation from the input mesh will be introduced. The extent of such deviation depends from the curvature and the sampling density, but will in general be small and therefore likely not relevant for practical applications. With our hybrid operator users can balance mesh growth with geometric fidelity by choosing the planarity threshold above which the switch from vertex split to edge split occurs.

%% file: 04_results.tex
\begin{figure}
	\centering
	\includegraphics[width=.91\linewidth]{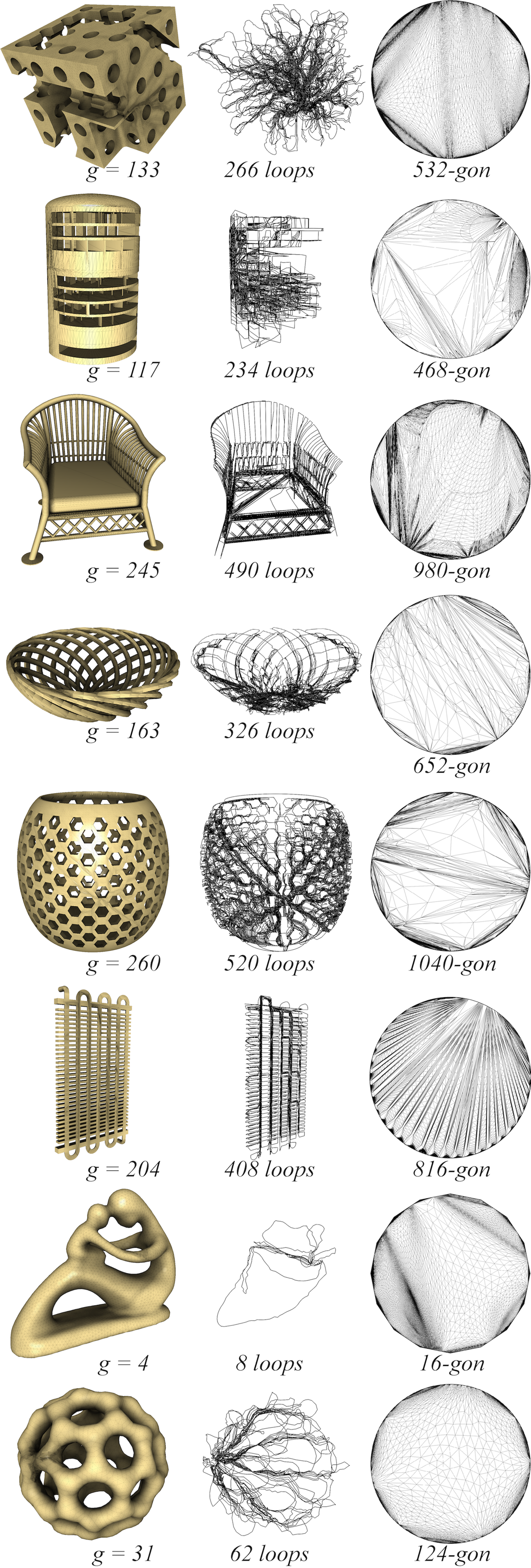}
	\caption{Gallery of results produced with the hybrid splitting scheme. From left to right: refined mesh, system of loops, canonical polygonal schema.}
	\label{fig:mosaic}
\end{figure}	

% \usepackage{multirow}
% \usepackage{graphicx}
\begin{table*}[]
	\centering
	\resizebox{\textwidth}{!}{%
		\begin{tabular}{lcr|c|r|c|c|r|c|c|r|c|c|l|c|}
			\cline{4-15}
			& \multicolumn{1}{l}{} & \multicolumn{1}{l|}{} & \multicolumn{3}{c|}{} & \multicolumn{3}{c|}{} & \multicolumn{6}{c|}{} \\
			&  & \multicolumn{1}{c|}{} & \multicolumn{3}{c|}{} & \multicolumn{3}{c|}{} & \multicolumn{6}{c|}{} \\ \cline{1-3}
			\multicolumn{1}{|l|}{\textbf{}} & \multicolumn{1}{l|}{\textbf{}} & \multicolumn{1}{l|}{\textbf{}} & \multicolumn{3}{c|}{\multirow{-3}{*}{\textbf{\begin{tabular}[c]{@{}c@{}}Edge\\ Split\end{tabular}}}} & \multicolumn{3}{c|}{\multirow{-3}{*}{\textbf{\begin{tabular}[c]{@{}c@{}}Vert\\ Split\end{tabular}}}} & \multicolumn{6}{c|}{\multirow{-3}{*}{\textbf{\begin{tabular}[c]{@{}c@{}}Hybrid\\ Split\end{tabular}}}} \\
			\multicolumn{1}{|l|}{\textbf{Model}} & \multicolumn{1}{c|}{\textbf{\#V/\#T}} & \multicolumn{1}{c|}{\textbf{genus}} & \textbf{\#V/\#T} & \multicolumn{1}{c|}{\textbf{\%}} & \multicolumn{1}{c|}{\textbf{Val.}} & \textbf{\#V/\#T} & \multicolumn{1}{c|}{\textbf{\%}} & \multicolumn{1}{c|}{\textbf{Val.}} & \textbf{\#V/\#T} & \multicolumn{1}{c|}{\textbf{\%}} & \multicolumn{1}{c|}{\textbf{Val.}} & \multicolumn{1}{l|}{\textbf{Cop.}} & \multicolumn{1}{c|}{\textbf{V/E \%}} & \textbf{H} \\ \hline
			\multicolumn{1}{|l|}{Bamboo Basket} & \multicolumn{1}{c|}{8K/16K} & 163 & \multicolumn{3}{c|}{\cellcolor[HTML]{FFCCC9}out of memory (\textgreater{}16GB)} & 26K/53K & 237\% & 114/17.6 & 337K/675K & 4229\% & 43/14.3 & $5^\circ$ & 64\%/36\% & $2\text{e-}3/3\text{e-}5$ \\ \hline
			\multicolumn{1}{|l|}{Buckyball} & \multicolumn{1}{c|}{5K/11K} & 31 & 16K/32K & 202\% & 22/9.5 & 7K/14K & 35\% & 16/8.5 & 13K/26K & 145\% & 22/9.1 & $5^\circ$ & 40\%/60\% & $1\text{e-}3/7\text{e-}6$ \\ \hline
			\multicolumn{1}{|l|}{Chair} & \multicolumn{1}{c|}{47K/94K} & 245 & \multicolumn{3}{c|}{\cellcolor[HTML]{FFCCC9}out of memory (\textgreater{}16GB)} & 98K/197K & 109\% & 207/28.4 & 2.7M/5.4M & 5772\% & 156/24.9 & $5^\circ$ & 70\%/30\% & $1\text{e-}3/4\text{e-}6$ \\ \hline
			%			\multicolumn{1}{|l|}{Dehydrogenase} & \multicolumn{1}{c|}{91K/185K} & 607 & \multicolumn{3}{c|}{\cellcolor[HTML]{FFCCC9}out of memory(\textgreater{}16GB)} & 246K/494K & 169\% & 257/27.2 & \multicolumn{6}{c|}{\cellcolor[HTML]{FFCCC9}out of memory(\textgreater{}16GB)} \\ \hline
			\multicolumn{1}{|l|}{Eight} & \multicolumn{1}{c|}{0.75K/1.5K} & 2 & 0.78K/1.5K & 2\% & 9/6.9 & 0.75/1.5K & 1\% & 9/6.9 & 0.78K/1.5K & 1\% & 9/6.9 & $5^\circ$ & 22\%/78\% & $1\text{e-}3/3\text{e-}6$ \\ \hline
			\multicolumn{1}{|l|}{Fertility} & \multicolumn{1}{c|}{6K/12K} & 4 & 6K/12K & 4\% & 11/9.2 & 6K/12K & 2\% & 11/8 & 6K/12K & 4\% & 11/9.1 & $5^\circ$ & 11\%/89\% & $6\text{e-}4/1\text{e-}6$ \\ \hline
			\multicolumn{1}{|l|}{Hexbowl} & \multicolumn{1}{c|}{34K/70K} & 260 & \multicolumn{3}{c|}{\cellcolor[HTML]{FFCCC9}out of memory(\textgreater{}16GB)} & 110K/220K & 219\% & 163/13.1 & 875K/1.7M & 2450\% & 164/12.7 & $5^\circ$ & 60\%/40\% & $5\text{e-}4/1\text{e-}5$ \\ \hline
			\multicolumn{1}{|l|}{Nasty Cheese} & \multicolumn{1}{c|}{30K/61K} & 133 & 1.5M/3M & 4902\% & 49/11.1 & 48K/97K & 60\% & 93/11.5 & 177K/355K & 490\% & 62/11.8 & $5^\circ$ & 63\%/37\% & $7\text{e-}4/7\text{e-}6$ \\ \hline
			\multicolumn{1}{|l|}{Polycube} & \multicolumn{1}{c|}{1K/2K} & 50 & 47K/93K & 57016\% & 39/8.2 & 10K/21K & 1189\% & 29/7.9 & 10K/21K & 1197\% & 182/28.1 & $5^\circ$ & 99\%/1\% & $0/0$ \\ \hline
			\multicolumn{1}{|l|}{Radiator} & \multicolumn{1}{c|}{33K/68K} & 204 & \multicolumn{3}{c|}{\cellcolor[HTML]{FFCCC9}out of memory(\textgreater{}16GB)} & 65K/130K & 92\% & 578/123.9 & 2.1M/4.2M & 6155\% & 316/78.3 & $5^\circ$ & 81\%/19\% & $7\text{e-}5/2\text{e-}6$ \\ \hline
			\multicolumn{1}{|l|}{Thingi 1764652} & \multicolumn{1}{c|}{4.2K/9K} & 117 & 32K/65K & 7627\% & 32/15.1 & 11K/22K & 153\% & 45/13.1 & 54K/109K & 1179\% & 43/14.8 & $5^\circ$ & 60\%/40\% & $6\text{e-}4/2\text{e-}6$ \\ \hline
		\end{tabular}%
	}
	\vspace{1em}
	\caption{Numerical results for the three splitting operators. We report: number of vertices and triangles in input (\textbf{\#V/\#T}); mesh growth, measured as the percentage of newly inserted vertices in the mesh w.r.t. the initial vertex count (\textbf{\%}); max/avg valence of the merging vertices processed during refinement (\textbf{Val.}). Additionally, for the hybrid split operator we report the coplanarity threshold we used (\textbf{Cop.}), the percentage of vertex and edge splits executed (\textbf{V/E \%}), and max/avg Hausdorff distance from the input mesh (\textbf{H}).}
	\label{tab:results}
\end{table*}

\section{Results and Discussion}
\label{sec:results}
In this section we analyze the performances of the refinement strategies previously presented. Since we are mostly interested in the scalability of these operators, we focused our analysis on high genus meshes, which we mostly gathered from the Thingi10K~\cite{Thingi10K} dataset. For completeness, a few meshes with lower genus have also been considered (e.g. Eight, Fertility).
%Since in practical applications both mesh size and geometric fidelity are important, we analyze both aspects. 
%We considered a variety of different meshes, ranging from low genus ($g < 10$, e.g. Eight, Fertility), to high ($g<100$, e.g. Buckyball) and very high genus ($g > 100$, e.g. Chair, Hexbowl). Challenging models were mostly selected from the Thingi10K~\cite{Thingi10K} dataset, where we looked for closed orientable manifolds with high genus.
%Both smooth surfaces and CAD-like models with sharp creases have been considered. 
Our experimental setup is as follows: for each model we first compute a generic system of loops with~\cite{erickson2005greedy}. To reduce the computational effort, rather than computing the globally shortest system of loops we randomly pick a mesh vertex and compute the shortest system centered at it. We then apply the three refinement algorithms to detach all loops except at their basis, producing three alternative cut graphs that admit a canonical polygonal schema (Figure~\ref{fig:mosaic}).

In Table~\ref{tab:results} we report numerical results. For each refinement strategy we report mesh size before and after refinement, and the growth rate, measured in percentage w.r.t. the initial number of vertices. Since for the edge split and the hybrid strategies the number of elements incident to a merging vertex is deeply connected with the amount of new elements introduced in the mesh, we also report minimum and average valence of the merging vertices processed during refinement. Note that the maximum valence reported is always lower than $4g$ (with $g$ being the genus), which is the valence of the vertex at which the system of loops is centered. 
For the hybrid scheme, we also report the amount of vertex and edge splits executed, the coplanarity threshold we used to choose between them, as well as the maximum and average Hausdorff distances from the input manifold, computed with Metro~\cite{cignoni1998metro}. 
%For the coplanarity,  We used this threshold as default parameter for all the results shown in the paper, unless stated differently.

Looking at numbers, it becomes very clear that for low genus meshes there is no significant difference between the three refinement operators we presented. This is not surprising, in fact low genus manifolds require a small amount of splits, which do not impact mesh size whatever strategy is used. For meshes with growing genus the weaknesses of the edge split operator become evident, and manifolds grow up to several orders of magnitude with respect to their original size. In four cases our reference implementation was not even able to complete the refinement, as the mesh was bigger than the memory at disposal in our testing hardware (16GB of RAM). Conversely, the vertex split operator always provides minimal mesh refinement, exhibiting a mesh growth that is always lower than 250\% except for one mesh. The hybrid refinement stays in between, and the amount of mesh growth it produces closely relates with the geometry of the manifold. For largely planar meshes like the polycube, its performances are almost equivalent to the vertex split operator (only 1\% of edge splits performed), whereas for smooth geometries the amount of edge splits may grow up to 89\% (for Fertility, the worst result in this regard). Note that these numbers depend from the coplanarity threshold of choice. Smaller angles would increase the amount of edge splits, whereas larger angles would reduce it, possibly at the cost of a bigger deviation from the input geometry. We empirically observed that a threshold of 5 degrees provides a good tradeoff between the amount of vertex splits executed and the deviation from the input mesh. With this value the average Hausdorff distance was always below 4e-5. Users can intuitively play with this parameter, trading mesh size for geometric accuracy.
%while for low valence manifolds there is almost no difference between all the split operators, as soon as the genus grows the meshes produced with edge split are much
% bigger, even orders of magnitude bigger than the ones produced with the vertex split or with the hybrid split strategy. 

Taking a closer look at why the disparity between the edge split and the vertex split is so big, one plausible explanation is that the edge split introduces, for each merging vertex processed, an amount of new mesh elements that scales linearly with the size of the triangle fan being split (Section~\ref{sec:edge_split}). To this end, it should be noted that even though the average vertex valence for a triangle mesh is 6, the connectivity generated by these refinement operators produces vertices with bigger valence, especially for high genus meshes where multiple loops snap to the same path. To make an example, if at a merging vertex there are three incoming edges, each one carrying 10 loops, after detachment the outgoing vertex adjacent to it will have valence 32 (30 plus 2 incident edges in the original one ring the merging vertex). To fully detach all loops, one similar vertex will be produced as many times as the number of discrete steps between the initial position of the merging vertex and the root of the system of loops.
%each time the merging vertex will be moved one step forward towards the origin of the system of loops, a new vertex with same valence will be produced. 
Conversely, the vertex split operator only adds two new triangles and one new vertex for each loop being detached, and is therefore not affected by the valence of the merging vertices, which is in average higher for this scheme (Table~\ref{tab:results}).
%\cino{introduce experiment with polycubes here...}

\begin{figure}
	\centering
	\includegraphics[width=.9\linewidth]{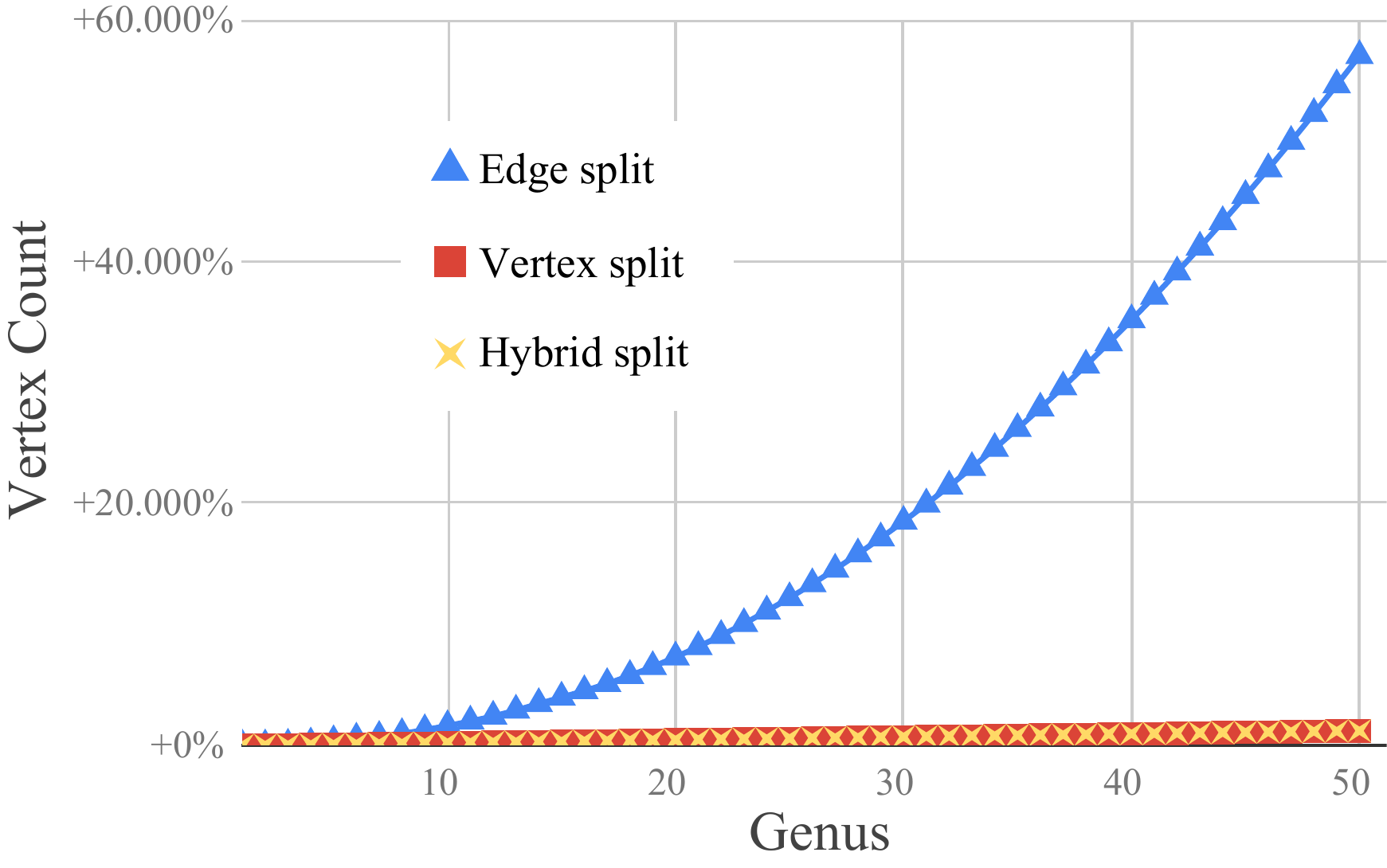}
	\caption{Mesh growth obtained by using the edge split, vertex split, and the hybrid split operators to refine a greedy homotopy basis of a sequence of polycubes with increasing genus.
		%Growth is measured as the ratio between the amount of newly inserted vertices and the number of vertices in the original mesh, and is given in percentage. 
		The edge split operator shows superquadratic growth ($57K\%$ new vertices for a polycube with genus 50). The vertex and hybrid split operators exhibit a similar linear growth, introducing $1188\%$ and $1197\%$ new vertices for the polycube with highest genus, respectively.}
	\label{fig:growth}
\end{figure}	

In Figure~\ref{fig:growth} we study the growth rate of the three splitting operators on a sequence of meshes with increasing genus. For this experiment, we considered as base mesh a genus one polycube like the one showed in Figure~\ref{fig:pcmap}, and produced a sequence of polycubes concatenating multiple occurrences of it. Overall, we produced 50 meshes with genus going from 1 to 50. As shown in the plot, the edge split operator has the worst performances, exhibiting a super quadratic growth in the number of vertices. After detaching all the loops in the system, the mesh of the torus with genus 50 contains $57K\%$ more vertices than its original version. Conversely, the vertex split operator exhibits a linear growth in mesh size, having a maximal growth in the number of vertices equal to $1188\%$. Interestingly, also the hybrid scheme has the same asymptotic behaviour, and has a maximal growth in the number of vertices equal to $1188\%$. As already pointed out before, note that the performances of the hybrid scheme depend from both topology and geometry, and may therefore be different for manifolds with different embedding. Nevertheless, polycubes are an interesting case of study because of their practical relevance: maps between a shape and a polycube abstraction of it are at the base of many applications, such as texturing, hexmeshing and spline fitting~\cite{LVSGS13,li2008globally}.
%it is interesting to study polycubes because they are prominent shapes in the context of cross maps between manifolds, and are widely used in a variety of applications, including .
%being not a purely topological scheme, results also depend from the geometry of the mesh, but we tend to believe that the behviour shown in this experiment will replicate in most of the practically relevant scenarios.

%\cino{discuss results in the table, mentioning high valence meshes, both smooth and not, and the importance of vertex valence and bla bla bla}
%\cino{TODO2: show a variety of high genus results, such as nasty cheese, somethign from thingi10k.}
%\cino{TODO3: show a nice polycube map made with polycut}

%% file: 05_conclusions.tex
\section{Conclusions and future works}
\label{sec:conclusions}
%Mapping a manifold with genus $g$ to a regular $4g-$gon is important in computer graphics and engineering because it allows to easily initialize a cross parameterization between any two shapes in the same homotopy group. For this map to be practically useful the cut graph should be shortest, but this latter condition makes the problem NP-hard. An interesting alternative consists in computing in polynomial time a shortest system of loops that possibly overlap at some mesh edge -- with~\cite{erickson2005greedy} -- and then detach such loops in post processing via mesh refinement. 

%This problem was already taken into consideration in~\cite{li2008globally} and~\cite{vegter1990computational}, where a refinement scheme based on the edge split operator was proposed. 

We showed that detaching cycles in a system of loops using the edge split operator
 %based on the edge split operator 
 triggers a mesh growth that explodes with genus, producing overly big meshes with little practical usefulness. In alternative we propose two novel refinement operators. The first one is based on the vertex split, and outperforms methods based on the edge split by introducing a minimal amount of elements in the mesh. Despite optimal from a topological point of view, this scheme may introduce deviation from the input surface. The second alternative addresses this limitation, and proposes to use as many vertex splits as possible, switching to the costly edge split only when significative surface deviation occurs.
 %purely topological, and uses the vertex split operator instead of the edge split, and despite showing a much better scalability it may introduce geometric deviation from the input mesh. The second one combines the positive aspects of the edge and the vertex split operators, using as many vertex splits as possible and switching to the costly edge split only when geometric deviation would be excessive. 
An intuitive parameter that measures the local planarity of the mesh allows users to trade between mesh size and geometric fidelity. 
In the technical part we also describe two simple heuristics to evaluate local planarity and to robustly position new mesh vertices. Although these methods worked fine in all our experiments, depending on the needs they could be easily substituted with more accurate (or faster) alternatives.

%study a variety of alternative refinement schemes, highlighting pros and cons of each scheme. In particular, we showed that the edge split strategy proposed in~\cite{li2008globally} induces a superquadratic mesh growth with respect to mesh genus, whereas refinement schemes based on the vertex split operator have a better (i.e. linear) scalability but may deviate from the input surface. We ultimately propose a hybrid refinement scheme that combines these two operators, thus minimizing the mesh growth and at the same time avoiding excessive deviation from the input geometry. An intuitive parameter that measures the local planarity of the mesh allows the user to trade between mesh size and geometric fidelity.

We support our claims with a variety of results, obtained on discrete manifolds that span from low to very high genus, and from smooth to CAD-like shapes. The proposed algorithms are based on well established local operators for simplicial meshes. These operators are already implemented in many geometry processing toolkits, making our results easy to reproduce. Nevertheless, we release a reference implementation of all the splitting methods presented in this paper (included the basic edge split strategy) inside Cinolib~\cite{cinolib}. 

%The work presented in this article should be seen as a building block of a robust pipeline to generate cross maps between shapes of any complexity. In perspective, we believe that coupling~\cite{erickson2005greedy} with our hybrid refinement scheme to cut a mesh open with minimal growth and high geometric fidelity, could be seen as 

Despite conceptually simple, we believe that this work makes one step forward towards the robust and computationally affordable generation of cross maps between complex shapes. Interesting results have already been presented for disk-like topologies~\cite{schmidt2019distortion}, and we expect more and more papers to come in future years.
%that the ability of our hybrid refinement scheme to cut a mesh open with minimal growth has the potential to foster new research in the field of surface maps, where
In the same spirit of recent works for the robust computation of planar maps, which start with Tutte's embedding and then cure distortion~\cite{rabinovich2017scalable,smith2015bijective,liu2018progressive}, we foresee a similar pipeline for cross maps between shapes, where manifolds are first cross mapped via their canonical schema, and then the polygon is evolved to minimize distortion. Note that this problem is much harder: partly because distortion minimization should consider the composition of two maps that overlap to one another, but more importantly because there are $4g$ alternative ways to overlap two $4g-$gons (i.e., which handle maps to which?), which makes it a problem with mixed discrete and continuous degrees of freedom for which, to the best of our knowledge, no effective solution is available in literature. Finally, we also plan to study the suitability of our cut graphs as an alternative to the ones described in~\cite{campen2019seamless} for the computation of global seamless parameterizations. 

%In particular, we are working on a pipeline where shapes of any genus are first cut open and mapped to their canonical polygonal schema to robustly initialize a discrete map; then, the map is further improved minimizing geometric distortion. Similar approaches have already been used for the robust computation of planar maps~\cite{rabinovich2017scalable},\cite{ liu2018progressive},\cite{smith2015bijective},\cite{shtengel2017geometric},\cite{claici2017isometry}. 
%The problem we describe here is more complex, and many questions remain open, mainly because 
%there are $4g$ alternative ways to overlap two $4g-$gons (i.e., which handle maps to which?), and distortion minimization should consider the composition of two planar maps that overlap to one another, and not a single map to the plane. 
%In future years, we expect a lot of research in the field from our community.